\documentstyle[emulateapj]{article}
\submitted{accepted to Ap.J. Letters}
\def\msun{{\rm\,M_\odot}}
\def\yr{{\rm\,yr}}

\def\gm{{\rm\,g}}
\def\cm{{\rm\,cm}}
\def\sec{{\rm\,s}}
\def\erg{{\rm\,erg}}
\def\kev{{\rm\,keV}}
\def\ev{{\rm\,eV}}
\def\K{{\rm\,K}}
\def\spose#1{\hbox to 0pt{#1\hss}}
\def\lta{\mathrel{\spose{\lower 3pt\hbox{$\mathchar''218$}}
     \raise 2.0pt\hbox{$\mathchar''13C$}}}
\def\gta{\mathrel{\spose{\lower 3pt\hbox{$\mathchar''218$}}
     \raise 2.0pt\hbox{$\mathchar''13E$}}}
\lefthead{Gammie and Menou}
\righthead{Episodic Accretion in Dwarf Novae}
\begin{document}
\title{On the Origin of Episodic Accretion in Dwarf Novae}
\author{Charles F. Gammie and Kristen Menou \altaffilmark{1}}
\affil{Harvard-Smithsonian Center for Astrophysics, MS-51 \\
60 Garden St., Cambridge, MA 02138, USA.; cgammie@cfa.harvard.edu,
kmenou@cfa.harvard.edu}

\altaffiltext{1}{also DARC-CNRS, Observatoire de Paris-Meudon,
        92195 Meudon, C\'edex, France.}


\begin{abstract}

We show that dwarf nova disks in quiescence have rather low magnetic
Reynolds number, of order $10^3$.  Numerical simulations of magnetized
accretion disks suggest that under these conditions magnetohydrodynamic
turbulence and the associated angular momentum transport is sharply
reduced.  This could be the physical origin of episodic accretion in
dwarf nova disks.  If so, the standard disk instability model needs to
be revised.

\end{abstract}


\section{Introduction}

It is a commonplace that resistive diffusion of magnetic fields is
negligible in astrophysical plasmas.  The importance of resistive
diffusion can be measured via the magnetic Reynolds number
$$
Re_M \equiv {L V\over{\eta}},
$$
where $\eta$ is the resistivity, and $L$ and $V$ are a typical
lengthscale and velocity, respectively.  $Re_M$ is often large in
astrophysics because the length and velocity scales are large.  For
example, the decay time of magnetic fields in a sunspot, as estimated
in the introduction to \cite{par79}, is $300$ yrs, implying a magnetic
Reynolds number of order $10^7$.  The decay time of magnetic fields in
the center of the sun, or in the galactic disk, is even longer.  In
this letter we discuss an uncommon situation where $Re_M$ is small and
the decay time is short: dwarf nova disks in quiescence.

Dwarf novae (DN) are binary stellar systems in which a white dwarf
accretes matter through an accretion disk from a main-sequence
companion.  DN go through regular, but not periodic, outbursts in which
the luminosity of the disk increases sharply.  They exhibit a rich and
complex phenomenology, reviewed in detail by \cite{war95}.  DN are
often regarded as ``laboratories'' for the study of astrophysical disk
systems because they are bright, nearby, and have short characteristic
timescales; changes in our understanding of these systems therefore
have broad implications for other disk systems such as X-ray binaries,
young stellar objects, and active galactic nuclei.

The significance of low $Re_M$ in DN disks is connected to
recent advances in the theory of angular momentum transport in
disks.  \cite{bh91} showed that disks are linearly unstable in MHD,
thereby revealing the long-sought-for instigator of disk turbulence.
More recent work on convection and nonlinear hydrodynamic
instability makes it seem unlikely that these once promising
alternatives play any role in angular momentum transport in disks.
Simulations of the nonlinear development of the Balbus-Hawley
instability show that it transports angular momentum outward and acts as
a dynamo in the sense that it sustains a magnetic field in the presence
of dissipation  (see \cite{bh97} for a complete review).

Most significantly, simulations by \cite{hgb96} (hereafter HGB) show
that the nonlinear development of the instability is sensitive to the
presence of resistive diffusion.  A series of experiments in that paper
shows that at $Re_M \equiv L_z^2 \Omega/\eta = 10^4$ the magnetic field
is depressed, \footnote{Notice that a different definition of $Re_M$ is
used in HGB.} while at $Re_M = 2000$ the magnetic field dies away, as
does the angular momentum flux.  This suggests that at $Re_M \lesssim
10^4$ MHD turbulence and the associated angular momentum transport may
die away.  In this letter we show, using SS Cygni as a specific
example, that $Re_M$ is of order $10^3$ in dwarf nova disks in
quiescence.

In \S 2 we briefly summarize the standard disk instability model for
DN outbursts.  In \S 3 we estimate $Re_M$ for SS Cygni's disk in
quiescence.  In \S 4 we tentatively put forward a new limit cycle model
for DN outbursts.  We discuss the implications of these results in \S 5
and conclude in \S 6.

\section{The Disk Instability Model}

Historically, a variety of mechanisms have been proposed to explain
DN.  The disk instability model, invoking an intrinsic modulation of
the accretion rate in the disk, is now generally accepted as the
explanation of DN outbursts (see \cite{can93b} for a review and a
discussion of the historical development of the subject).  The essence
of the mechanism is that the disk cycles between two states, a large
accretion rate state (hot and mostly ionized) and a low accretion rate
state (cold and mostly neutral).  The disk cannot settle into a steady
intermediate state because that state is thermally unstable.

The simplest version of the disk instability model, in which the
angular momentum transport efficiency $\alpha$ (introduced by
\cite{ss73}) is held constant throughout the cycle, does not work,
however (e.g.  \cite{sma84}).  Detailed time-dependent calculations
show that, in order to obtain an outburst of sufficient amplitude,
$\alpha$ must vary between the hot and cold state.  Researchers have
been able to reproduce outburst light curves and mean times between
outbursts using $\alpha_{hot} \simeq 0.1$ and $\alpha_{cold} \simeq
0.01$.  Another oft discussed possibility is that $\alpha$ obeys a
scaling relation such as $\alpha = \alpha_0 (H/r)^q$, where $H$ is the
disk scale height and $r$ the local radius.  To our knowledge this
prescription has been used in the DN context only to study decay
properties of the outburst light curves.  Both these prescriptions for
varying $\alpha$ are phenomenologically motivated; while there exist
some {\it post hoc} theoretical justifications for varying $\alpha$, in
our view they are not strongly physically motivated.

\section{Resistivity in SS Cygni}

What then is $Re_M$ in dwarf nova disks?  For definiteness, we focus on
a nearby, well-studied system:  SS Cygni.  SS Cyg has a $1.2 \msun$
white dwarf accreting from a disk fed by a $0.7 \msun$ K5V companion, a
period of $6.6$hr, a distance from the primary to the L1 point of about
$6 \times 10^{10} \cm$, a mean interval between outbursts of $40$d, and
a characteristic decay time from outburst of $2.4$d.  The mean
accretion rate is not well known but is of order $10^{-9} \msun
\yr^{-1}$ (\cite{can93a}).  

We require a temperature and density to estimate the resistivity in SS
Cygni in quiescence.  These are only weakly constrained by
observations.  In the absence of direct measurements, we must turn to a
theoretical model for the disk evolution: the standard disk instability
model.  Future observations may provide better constraints on physical
conditions in the disk.

SS Cyg has been theoretically studied in detail using the standard
thermal limit cycle model (see \cite{can93a} for the most detailed
study to date).  We have run our own evolutionary models of SS Cyg to
estimate physical conditions in the quiescent disk, using a modern,
time-dependent, implicit, adaptive grid code (\cite{ham97}).   We
adopt the same parameters for SS Cyg as Cannizzo's standard model
($\alpha_{hot} = 0.1$, $\alpha_{cold} = 0.02$), and our calculated
$\Sigma$ and $T_c$ are similar to those in his standard model.  At a
radius of $2 \times 10^{10}\cm$ from the primary a typical surface
density in quiescence is $200 \gm \cm^{-2}$, and a typical {\it
central} temperature is $3000 \K$.  Then $\Omega = 4.5 \times 10^{-3}
\sec^{-1}$, $c_s = 3.5 \times 10^5 \cm\sec^{-1}$, $H = c_s/\Omega = 7.7
\times 10^7 \cm$, and $\rho \simeq \Sigma/(2 H) = 1.3 \times 10^{-6}
\gm \cm^{-3}$.  In LTE hydrogen is then predominantly molecular.  The
disk is marginally optically thick in a Rosseland mean sense.

The resistivity is given by
$$
\eta = {c^2 m_e \nu_c\over{4 \pi n_e e^2}},
$$
where $\nu_c$ is an effective collision frequency for electrons.  Since
electron-neutral collisions dominate, we take $\nu_c = n_n \sigma_{en}
v$, where $n_n$ is the neutral number density, $\sigma_{en}$ is the
electron-neutral momentum exchange cross section, and $v = \sqrt{128 k
T/(9\pi m_e)}$.  At $3000\K$ $\sigma_{en} = 1.3 \times 10^{-15} \cm^2$
(\cite{hay81}).  Then $\eta = 1.67 \times 10^4 n_n/n_e$.  The electron
abundance at this temperature is determined primarily by ionization of
Na, and to a lesser extent, Ca and K (we assume solar abundances).  A
detailed solution for ionization equilibrium (we have used a code
kindly provided to us by Dr. P. Hoeflich, but a simplified calculation
involving Na, C, and K gives nearly identical results) gives $n_n =
3.73 \times 10^{17} \cm^{-3}$, and $n_e = 8.53 \times 10^{11}
\cm^{-3}$, so $\eta = 7.29 \times 10^{9} \cm^{2} \sec^{-1}$.

The natural length and velocity scales in disks are $H$ and $c_s$,
respectively.  The disk magnetic Reynolds number is then $Re_M \equiv
c_s H/\eta$.  Using the resistivity just calculated, $Re_M = 3670$.
This is below the value at which MHD turbulence is depressed in the
simulations of HGB.  Similarly low values of $Re_M$ are obtained at
other radii in our model.  Of course, this $Re_M$ was obtained assuming
$\alpha_{cold} = 0.02$.  At lower $\alpha$ the disk will be cooler and
denser, and then $Re_M$ will be smaller since $\eta$ is very sensitive
to temperature.  Since $\alpha$ depends on the resistivity as well,
there could be a runaway decay of MHD turbulence.  All this suggests
that, at least in SS Cygni, angular momentum transport may die away
once the disk goes into quiescence.

Before going on, let us briefly consider two other issues.  First, does
ambipolar diffusion play any role?  The relative importance of
ambipolar diffusion and resistivity is controlled by the product of the
electron and ion Hall parameters, $DR = \omega_e
\omega_i/(\nu_{en}\nu_{in})$, where $\omega_{e,i}$ are the ion and
electron cyclotron frequencies and $\nu_{in}$ is the ion-neutral
collision frequency;  when $DR > 1$ ambipolar diffusion dominates.
Assuming equipartition magnetic fields at our fiducial point in the SS
Cyg disk, we find $DR = 0.02$.  Since the field is likely to be weaker
than this, resistivity dominates ambipolar diffusion.

Second, can nonthermal ionization save the day?  Here the chief concern
is X-rays; dwarf novae typically have $L_x \simeq 10^{30}\erg
\sec^{-1}$, but SS Cyg is particularly bright in hard X-rays in
quiescence, with $L_x (1-37 \kev) \simeq 1.5 \times 10^{32} \erg
\sec^{-1}$ (\cite{yio92}).  Only photons with $E \gtrsim 10 \kev$ will
penetrate the disk.  These will Compton scatter and diffuse downward
into a layer of thickness $\simeq 30 \gm \cm^{-2}$ corresponding to an
effective optical depth of $1$ (\cite{gni97}, \cite{ig97}).  The
intercepted flux per unit area of the disk is $f_x \simeq L_x
(H/r)/(4\pi r^2)$.  Suppose this flux diffuses into a layer of
thickness $H$ and produces 1 ion per $E_i = 36 \ev$.  Then the volume
ionization rate is $\xi = L_x /(4 \pi r^3 E_i) \cm^{-3}$.  Balancing
this against dissociative recombination at a rate $n_e^2 \times 8.7
\times 10^{-6} T^{-1/2} \cm^{-3}$, we find $n_e = 1.1 \times 10^9
\cm^{-3}$.  Since this is much less than the LTE electron number
density we can neglect X-ray ionization in the bulk of the disk.  It is
possible, however, that a thin layer on the surface of the disk will be
ionized by X-rays, and accretion will proceed in that layer in a manner
similar to that described by \cite{gam96} in the context of
protoplanetary disks.  A thin partially ionized layer could explain
eclipse maps of quiescent dwarf novae (\cite{whv92}, \cite{hor93}),
which are interpreted as showing emission from a warm optically thin
disk.

Suppose that MHD turbulence dies away in SS Cyg's disk in quiescence.  It
is possible that some weaker residual transport process will be present
in the disk.  Let us consider some of the possibilities.  Convective
turbulence is ruled out for at least two reasons: the disk is only
marginally optically thick, so convection will not be present, and even
it were present, recent studies show that convective turbulence
produces {\it inward} angular momentum transport (\cite{rg92},
\cite{sb96}, \cite{cab96}).  Turbulence due to nonlinear hydrodynamic
instability also seems unlikely in light of recent analytic and
numerical work (\cite{bhs96}, \cite{hb97}) which shows that nonlinear
instability is simply not present in a Keplerian disk.  Gravitational
instability is ruled out, since for reasonable values of the surface
density and of the ``floor'' temperature, set by irradiation of the
disk by the secondary, hot spot, and primary, the disk has $Q \gg 1$.
The remaining possibilities involve a spiral wave or shock, driven
either by the companion's tidal field (\cite{smh86}; \cite{spr87}), or
by a global linear instability (\cite{pp84}, \cite{ggn86}).

The role of tidally driven spiral shocks is controversial 
(\cite{rs93}, \cite{spl94}; see also the claimed detection of spiral
waves in IP Peg in outburst by \cite{shh97}).  In our view, Savonije et
al.  have shown that a disk with small $H/r$ and fixed mass couples
only weakly to the tidal potential, and suffers no global instability.
They have not, however, considered the interaction of the disk with the
mass transfer stream.  Since the incoming material has approximately
constant specific angular momentum, which will be conserved in the
absence of turbulent diffusion, it will accumulate in a partially
pressure supported ring.  This ring or torus may then suffer the
\cite{pp84} instability, or an allied global instability, as suggested
by \cite{rs93}.

\section{A New Limit Cycle?}

The possibility of global hydrodynamic instabilities in DN disks leads
us to tentatively propose a variant of the classical limit-cycle model
for DN outbursts.  In the hot state the evolution proceeds as usual
until accretion causes the surface density to drop below the point
where the disk can maintain the hot solution.  The disk then cools,
recombines, and forms molecular hydrogen.  If the magnetic Reynolds
number is low enough, then MHD turbulence will decay away.  The disk
then enters quiescence (the viscosity could even be as low as the
molecular value, if no other angular momentum transport process is at
work), and new material accumulates in a partially pressure-supported
ring in the outer disk.  Eventually this ring suffers a global
hydrodynamic instability.  The resulting shocks raise the central
temperature of the disk, and hence $Re_M$, above a critical value.  MHD
turbulence sets in, raising the temperature still further and providing
significant angular momentum transport.  An outburst results.  This
idea is illustrated in Figure 1.

It remains to be seen whether this scenario can be integrated into a
successful time-dependent model for the outburst.  We suspect that at
least one additional ingredient is required, and that is a modification
of the classical diffusion equation for surface density in a viscous
disk (e.g. \cite{pr81}).  An important physical assumption underlying
this equation is that the local stress adjusts instantaneously to a
value specified by the current surface density and temperature.
Simulations of HGB and \cite{bnst95} suggest that there is a relaxation
time of many orbital periods as the stress adjusts to the current
equilibrium value.  To our knowledge, no study of time-dependent
behavior in disks includes this effect.

\section{Discussion}

As a guide to where resistive effects become important in disks, we 
have calculated the temperature $T_{4}$, where $Re_M = 10^4$, as a
function of surface density and rotation frequency.  The results are
shown in Figure 2.  A fit accurate to $5\%$ is
$$
\log(T_{4}) = 3.867 + 0.038 \log(\Sigma) + 0.20 \log(\Omega)
        + 0.01 \log(\Omega)^2,
$$
where $\Sigma$ and $\Omega$ are expressed in $\gm\cm^{-2}$ and
$\sec^{-1}$ respectively.  Notice that, for a given value of $\alpha$
and for realistic opacities not all regions of this plot are
accessible.  In particular, disks in the upper right corner are
unlikely to be cool enough to reach $T_4$.

The above discussion is predicated on the idea that MHD turbulence in
disks decays below a critical $Re_M$ of order $10^4$.  This is
consistent with the evidence from numerical simulations, but cannot be
regarded as firmly established given the small number of simulations
that have been done to date and their low numerical resolution.
\footnote{The critical value of the resistivity, if any, must be
determined from nonlinear theory.  Linear theory says that stability is
recovered at a value of the resistivity that depends on the field
strength.  Absent an almost thermal external field, however, the field
strength in the disk is determined by the nonlinear outcome of the
instability itself.} One possible complication is that saturation may
depend on magnetic Prandtl number $Pr_M \equiv \nu/\eta$ rather than
simply on $Re_M$ (\cite{bat50}); here $\nu$ is the microscopic
viscosity.  The sense of the theoretical argument is that when
viscosity is larger than resistivity the field builds up, while in the
opposite limit the field dies away.  Counterintuitively, this argument
suggests that $\alpha$ should increase as $\nu$ increases.  There is
weak evidence for this from the simulations in that those with larger
{\it artificial} viscosity saturate with slightly larger $\alpha$.
Since microscopic viscosity in disks is small compared to the numerical
and artificial viscosities present in the simulations, there is the
possibility that MHD turbulence may decay at even higher temperatures
than we have indicated.  Clearly there are many unanswered questions
about the saturation of disk turbulence that might profitably be
addressed by future numerical simulations, with the promise of direct
application to astronomically interesting systems such as DN.

Several characteristics of the new limit cycle are worth mentioning
here.  First, our model bears some resemblance to that of \cite{alp96},
who speculated that MHD turbulence would cease as DN enter quiescence
because the magnetic field would become superthermal.  Our physical
mechanism is, however, different and less speculative.  We expect the
field to resistively decay, rather than being frozen in and dynamically
dominant.  Second, in our scenario the outburst is triggered rather
differently than in the disk instability model.  Assuming that
turbulence exists above a critical $Re_M$ outburst should begin when
the disk rises above a critical temperature, not when the disk becomes
thermally unstable.  Third, the details of the outburst will be
sensitive to metallicity, or at least the abundance of Na, because
Na$^+$ provides most of the free electrons at low temperatures.
Finally, preliminary calculations suggest that the model is applicable
to X-ray binaries, although disk irradiation could complicate matters
somewhat.

\section{Conclusion}

We have argued that MHD turbulence and the associated angular momentum
transport may die away in dwarf novae in quiescence.  If there are no
other significant sources of angular momentum transport in disks, then
this is the physical origin of episodic accretion in dwarf novae.  We
have proposed a scenario in which a global hydrodynamic instability
heats the outer disk, thereby raising the conductivity and initiating
the outburst.  This scenario is a modification of the standard disk
instability model that provides a physical explanation for episodic
accretion yet retains many of the standard model's most attractive
features.

\acknowledgments

We are grateful to S. Balbus, J.-M. Hameury, J. Hawley, S. Kenyon,
J.-P. Lasota, R. Narayan, E. Quataert, J. Raymond, H. Vanhala, and E.
Vishniac for helpful comments.  P. Hoeflich provided a code for
calculating electron abundances.  This work was supported by NASA grant
NAGW 5-2836.  KM was supported by an SAO Predoctoral Fellowship and a
French Higher Education Ministry grant.

\clearpage

\begin{figure}
\plotone{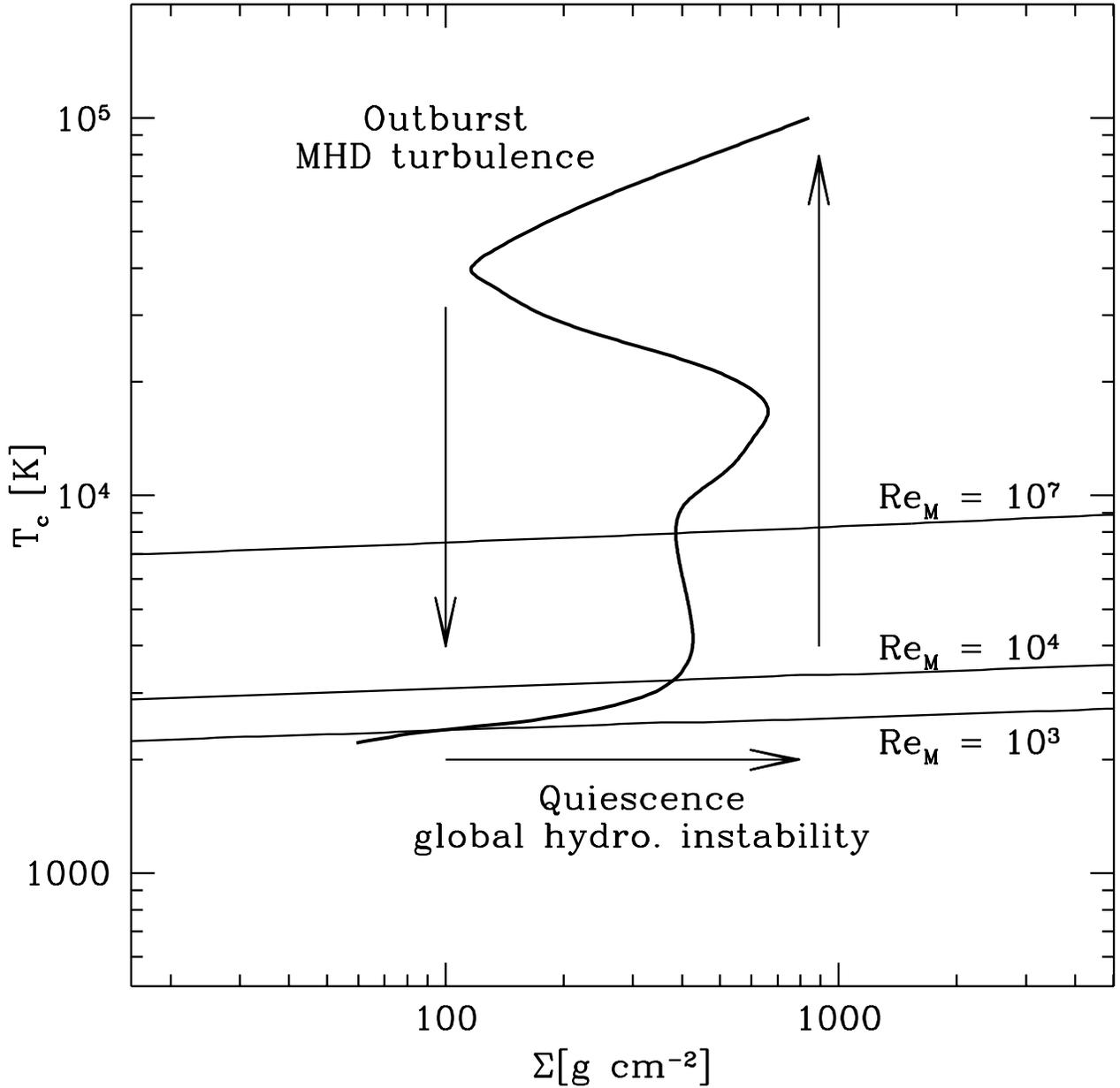}
\caption{
A schematic of the proposed outburst cycle.  The solid line
(``S-curve'') shows the thermal equilibrium solution for the disk in SS
Cygni at radius $r = 2 \times 10^{10}\cm$, using the standard
assumption of $\alpha = 0.1$ on the hot branch and $\alpha = 0.02$ on
the cold branch (the transition temperature is $2.5 \times 10^4\K$).
Our estimates show that when the disk ``falls off'' the hot branch, it
will pass through the line marked $Re_M = 10^4$, where MHD turbulence
dies away.  The disk is then passive, and is reactivated only by a
global hydrodynamic instability.
}
\end{figure}

\begin{figure}
\plotone{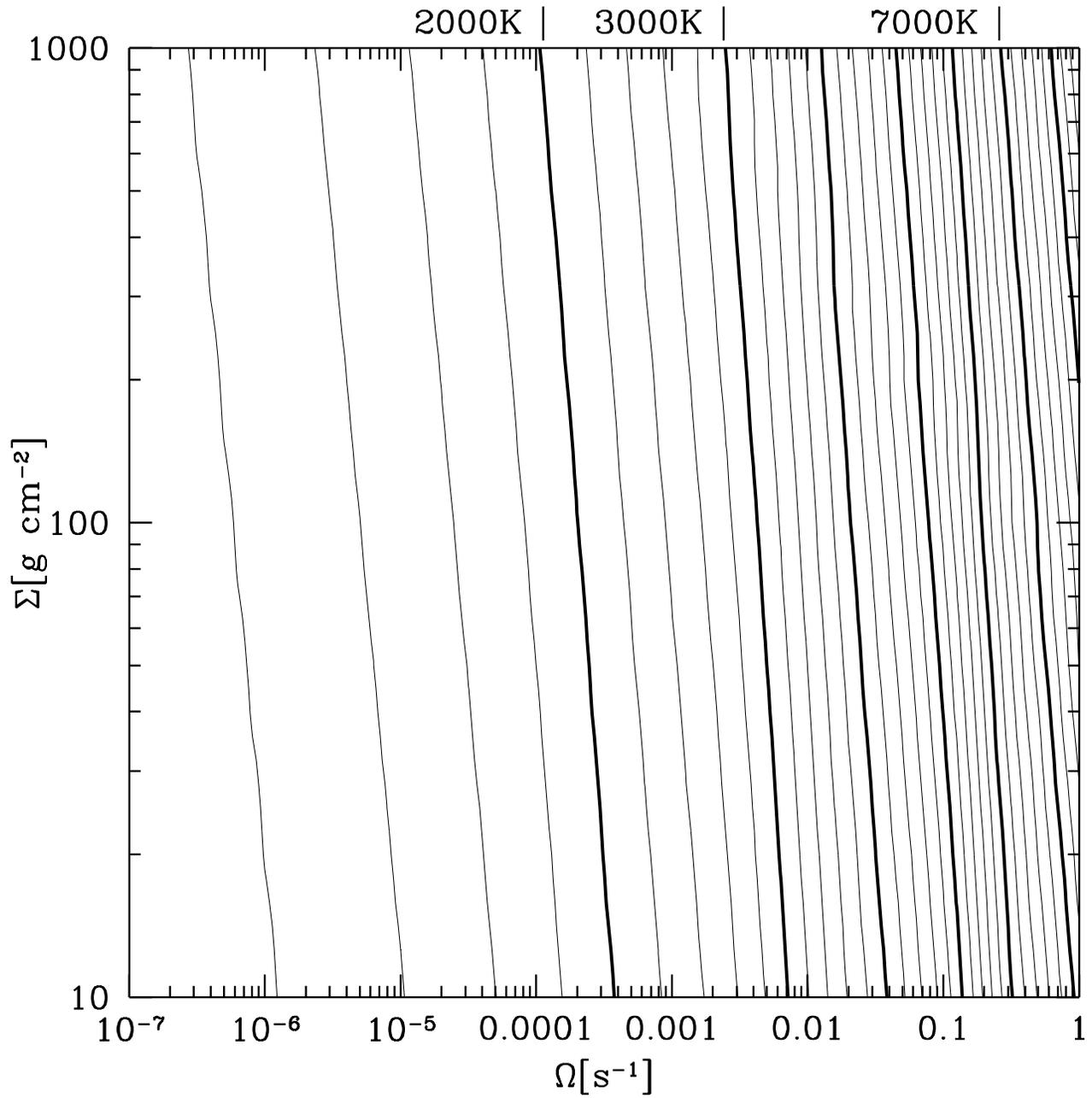}
\caption{
The temperature $T_{4}$ where $Re_M = 10^4$, as a function of rotation
frequency and local surface density.  The heavy solid lines are spaced
at intervals of $10^3\K$, the light lines at intervals of $200\K$.  
Recall that at $T > T_4$, $Re_M > 10^4$.
}
\end{figure}

\end{document}